\documentclass[sigconf,natbib,9pt]{acmart}

\copyrightyear{2018}
\acmYear{2018}
\setcopyright{acmlicensed}
\acmConference[SIGIR '18]{The 41st International ACM SIGIR Conference on Research and Development in Information Retrieval}{July 8--12, 2018}{Ann Arbor, MI, USA}
\acmBooktitle{SIGIR '18: The 41st International ACM SIGIR Conference on Research and Development in Information Retrieval, July 8--12, 2018, Ann Arbor, MI, USA}
\acmPrice{15.00}
\acmDOI{10.1145/3209978.3210079}
\acmISBN{978-1-4503-5657-2/18/07}

\usepackage[subtle, wordspacing=normal, tracking=normal, bibnotes, charwidths=normal, leading, indent, lists, paragraphs=normal, mathspacing=normal, bibliography=normal]{savetrees}
\pdfoutput=1

\usepackage{amsfonts}
\usepackage{siunitx}
\usepackage{multirow}
\usepackage{enumerate}
\usepackage{booktabs}
\usepackage{amssymb}
\usepackage{calc}
\usepackage{paralist}
\usepackage{subcaption}
\usepackage[labelfont=bf,textfont=bf,singlelinecheck=on]{subcaption}
\usepackage{tikz}
\usetikzlibrary{fit,positioning}
\usepackage{graphicx}
\usepackage[justification=centering]{caption}
\usepackage{color}
\usepackage{colortbl}
\usepackage{url}
\usepackage{mathtools}
\usepackage{algorithm}
\usepackage{algorithmic}
\usepackage[normalem]{ulem}
\usepackage{enumitem}
\usepackage{numprint}
\npthousandsep{,}
\npdecimalsign{.}
\usepackage[skip=0pt]{caption}
\usepackage{setspace}
\usepackage{color}
\usepackage{colortbl}
\usepackage{acronym}

\acrodef{AHNQS}{Attention-based Hierarchical Neural Query Suggestion}
\acrodef{RNN}{recurrent neural network}
\acrodef{NQS}{neural query suggestion}
\acrodef{HNQS}{hierarchical neural query suggestion}

\newcommand{\noop}[1]{}

\newcommand{\duwen}{\phantom{$^\blacktriangle$}}

\allowdisplaybreaks

\title{Attention-based Hierarchical Neural Query Suggestion}

\author{Wanyu Chen}
\orcid{}
\affiliation{%
\institution{Science and Technology on Information Systems Engineering Laboratory\\ National University of Defense Technology}
\city{Changsha}
\country{China}
}
\email{wanyuchen@nudt.edu.cn}

\author{Fei Cai}
\authornote{Corresponding author.}
\orcid{}
\affiliation{%
\institution{Science and Technology on Information Systems Engineering Laboratory\\ National University of Defense Technology}
\city{Changsha}
\country{China}
}
\email{caifei@nudt.edu.cn}

\author{Honghui Chen}
\orcid{}
\affiliation{%
\institution{Science and Technology on Information Systems Engineering Laboratory\\ National University of Defense Technology}
\city{Changsha}
\country{China}
}
\email{caifei@nudt.edu.cn}

\author{Maarten de Rijke}
\orcid{0000-0002-1086-0202}
\affiliation{%
\institution{Informatics Institute\\ University of Amsterdam}
\city{Amsterdam}
\city{The Netherlands}
}
\email{derijke@uva.nl}

\begin{document}

\begin{abstract}
Query suggestions help users of a search engine to refine their queries.
Previous work on query suggestion has mainly focused on incorporating directly observable features such as query co-occurrence and semantic similarity.
The structure of such features is often set manually, as a result of which hidden dependencies between queries and users may be ignored.
We propose an \ac{AHNQS} model that combines a hierarchical structure with a session-level neural network and a user-level neural network to model the short- and long-term search history of a user.
An attention mechanism is used to capture user preferences.
We quantify the improvements of \ac{AHNQS} over state-of-the-art \acl{RNN}-based query suggestion baselines on the AOL query log dataset, with improvements of up to 21.86\% and 22.99\% in terms of MRR@10 and Recall@10, respectively, over the state-of-the-art; improvements are especially large for short sessions.
\end{abstract}

%
%



\maketitle


\section{Introduction}
\label{Intro}
Modern search engines offer query suggestions to help users express information need effectively.
Previous work on query suggestion, such as probabilistic models and learning to rank techniques, mainly relys on features indicating dependencies between queries and users, such as clicks and dwell time~\citep{Relevant2003,PQSD2017}.
However, the structure of those dependencies is usually modeled manually. As a result, hidden relationships between queries and a user's behavior may be ignored. Consequently, \ac{RNN}\acused{RNN} based approaches have been proposed to tackle these challenges. A query log can be treated as sequential data that can be modeled to predict the next input query.
However, existing neural based methods only consider so-called current sessions (in which a query suggestion is being generated) as the search context for query suggestion~\citep{onal-neural-2018}.

We propose an \acf{AHNQS} model that applies a user attention mechanism inside a hierarchical neural structure for query suggestion.
The hierarchical structure contains two parts: a session-level \ac{RNN} and a user-level \ac{RNN}.
The session-level \ac{RNN} captures queries in the current session and is used to model the user's short-term search context to predict their next query.
The user-level \ac{RNN} captures the past search sessions for a given user and is applied to model the user's long-term search behavior to output a user state vector representing their preferences.
We use the hidden state of the session-level \ac{RNN} as the input to the user-level \ac{RNN}; the user state of the latter is then used to initialize the first hidden state of the next session-level \ac{RNN}.

In addition, we apply an attention mechanism inside the hierarchical structure that is meant to capture a user's preference towards different queries in a session. This addition is based on the assumption that different queries in the same session may express different aspects of the user's search intent~\citep{attention}, e.g., queries with subsequent click behavior are more likely to represent the user's information need than those without. An attention mechanism can automatically assign different weights for hidden states of the queries in the session-level \ac{RNN}. The attentive hidden states together compose the session state, which is used as input for the user-level \ac{RNN}.

We compare the performance of \ac{AHNQS} against a state-of-the-art query suggestion baseline and variants of \ac{RNN} based query suggestion methods using the AOL query log. 
In terms of query suggestion ranking accuracy we establish improvements of \ac{AHNQS} over the best baseline model of up to $21.86\%$ and $22.99\%$ in terms of MRR@10 and Recall@10, respectively.

Our contributions in this paper are:
\begin{inparaenum}[(1)]
   \item  We tackle the challenge of query suggestion in a novel way with neural network based method.
   \item  We propose \ac{AHNQS}, which adopts a hierarchical structure containing a user attention mechanism to better capture the user's search intent.
    \item We analyse the impact of session length on query suggestion performance and find that \ac{AHNQS}  consistently yields the best performance, especially with short search contexts.
\end{inparaenum}


\section{Approach}
\label{Approach}

\subsection{Session-level \acp{RNN} for query suggestion}
\label{section:NQS}

As in~\citep{RNN2015}, session-level \acp{RNN} are our starting point. Here in the neural based query suggestion model (NQS), queries in the current session are taken as sequential input and used to output the probability of being the next query for the query suggestion candidates.

\begin{figure}[t]
  \centering
   \includegraphics[width=0.37\textwidth]{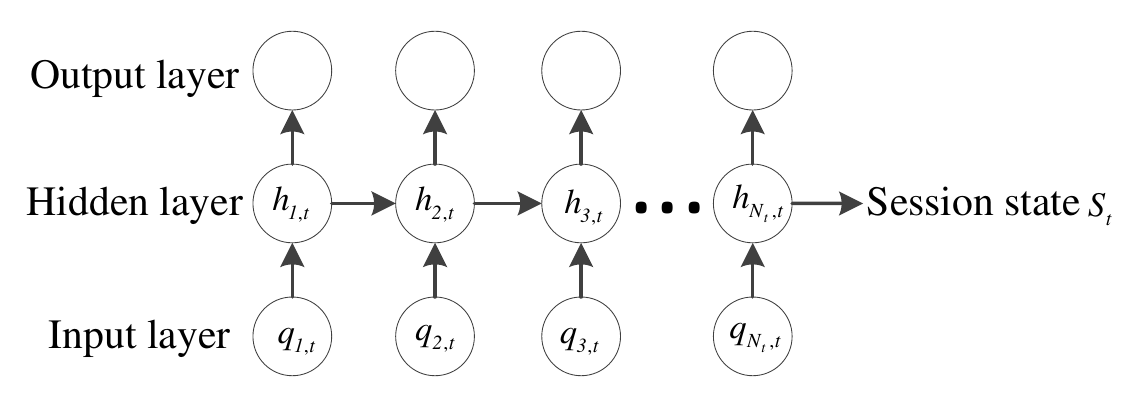}
      \caption{Structure of the \ac{NQS} model.}
\label{RNN-QS}
\end{figure}

Formally, we assume that a query session $\emph{Session}_t$ contains $N_t$ queries, denoted as  $\emph{Session}_t=(q_{1,t},q_{2,t},q_{3,t},\ldots ,q_{N_t,t})$.
%
%
As shown in Fig.~\ref{RNN-QS}, for generating the input vector of the network, we use a 1-of-$N$ encoding of $q_i$, i.e., the vector length equals the number of unique queries $V$
and only the coordinate corresponding to the $i$-th query is one, the others are zero.
We choose to use the Gated Recurrent Unit (GRU)~\citep{GRU} as our non-linear transformation.
The hidden state $h_{n}$ can be calculated by using the previous hidden state $h_{n-1}$ and the candidate update state $\widehat{h}_n$:
\begin{equation}
\label{equation1}
\textstyle
h_n=(1-u_n)h_{n-1}+u_n\widehat{h}_n,
\end{equation}
where the update gate $u_n$ can be generated by:
\begin{equation}
\label{equation2}
\textstyle
u_n=\sigma(\emph{$I_u$}q_{n,t}+\emph{$H_u$}h_{n-1}).
\end{equation}
The candidate update state $\widehat{h}_n$ is calculated by:
\begin{equation}
\label{equation3}
\textstyle
\widehat{h}_n=\tanh(\emph{$I$}q_{n,t}+\emph{$H$}(r_n \cdot h_{n-1})),
\end{equation}
where the reset gate $r_n$ is:
\begin{equation}
\label{equation4}
\textstyle
r_n=\sigma(I_rq_{n,t}+ H_r h_{n-1})
\end{equation}
where $I$, $I_u$, $I_r\in \mathbb{R}^{d_h\times V}$, $H$, $H_u$, $H_r \in \mathbb{R}^{d_h\times d_h}$, and $d_h$ is the number of dimension of the hidden state. The $H$ matrices are used to keep or forget the information in $h_{n-1}$.

We use $\emph{RNN}_\emph{session}$ and $\emph{RNN}_\emph{user}$ to denote the GRU function.
The final hidden state of a session-level \ac{RNN} is used to indicate the session state, $S_t=h_{N_t,t}$. The output of the session-level \ac{RNN} are the scores of query suggestion candidates being predicted as the next query:
$s_{n,t}=g(h_{n,t})$,
where $g(\cdot)$ is the active function of the output layer, which can be a softmax or $\tanh$ depending on the loss function of the neural network. We generate the query suggestion list in the test set according to the scores of query candidates.

We choose a pairwise loss function that forces positive query suggestion samples to be ranked higher than the negative ones.
Actually, there are several pairwise ranking loss functions, including cross-entropy and TOP1~\citep{2016session-based}.
In the field of recommender systems, TOP1 has proved to outperform others, so we set:
\begin{equation}
\label{equation6}
\textstyle
Loss=\frac{1}{N_S}\cdot \sum_{j=1}^{N_S} \sigma(s_{j,t}-s_{i,t})+\sigma(s_{j,t}^2),
\end{equation}
where $s_{j,t}$ and $s_{i,t}$ denote the score of a negative query candidate and a ground truth query, respectively.

\subsection{Hierarchical user-session \ac{RNN} for query suggestion}
\label{section:HNQS}
Clearly, the \ac{NQS} model only models the short-term search context.
Here, we model the long-term search behavior of a given user with a user-level RNN, producing the hierarchical NQS model (HNQS).
\begin{figure}[h]
  \centering
   \includegraphics[width=0.45\textwidth]{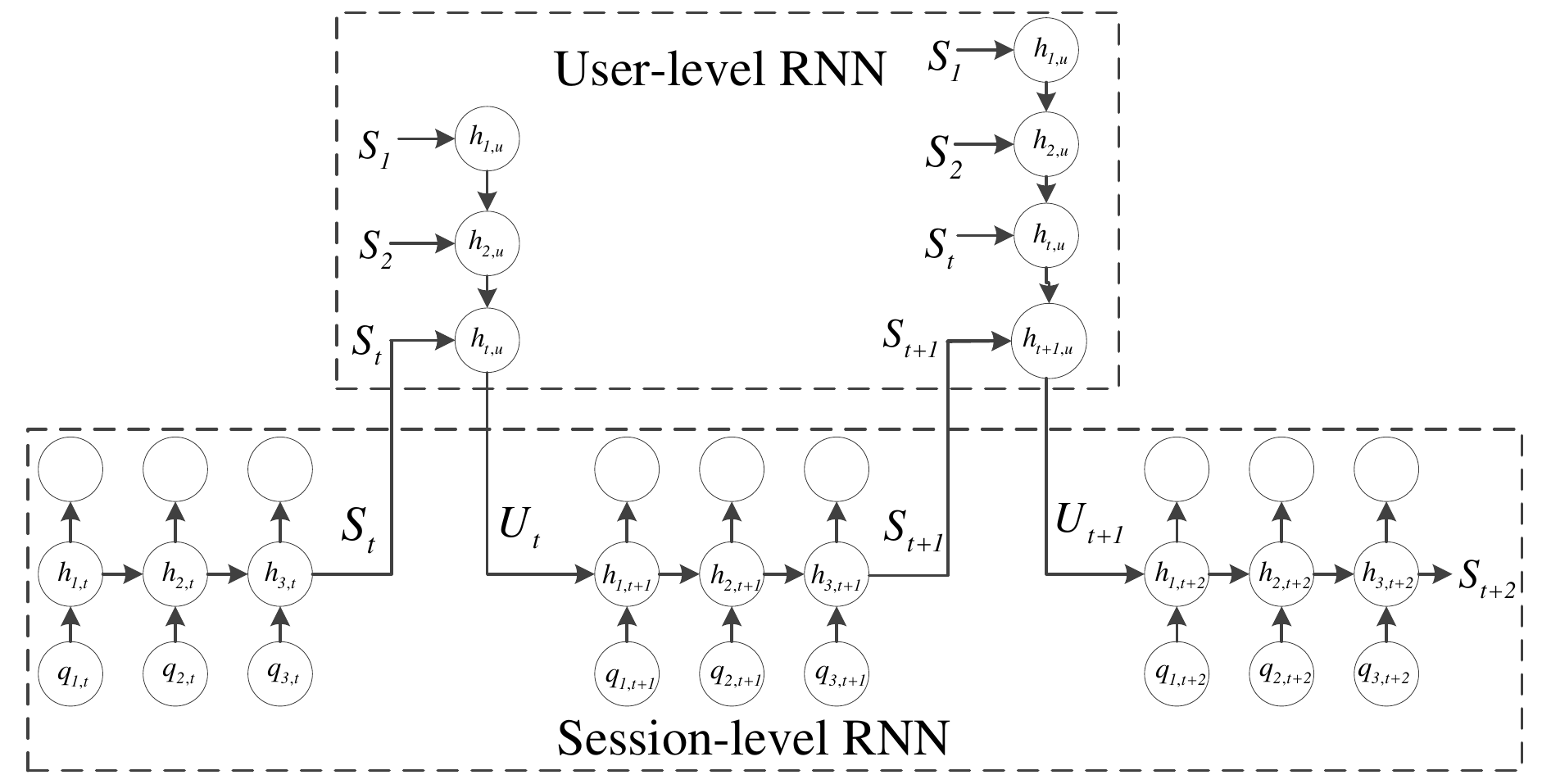}
   \caption{Structure of the \ac{HNQS} model.}
\label{HRNN-QS}
\end{figure}

We assume that a user $u$ has $N_u$ query sessions, $\emph{User}_u=(\emph{Session}_{1,u}$, $\emph{Session}_{2,u}, \ldots, \emph{Session}_{N_u,u})$. In a user-level \ac{RNN}, the input is the session state $S_t$ and the hidden state can be calculated as:
\begin{equation}
\label{equation7}
\textstyle
h_{n,u}=\emph{RNN}_\emph{user}(h_{n-1,u},S_{n,u}),
\end{equation}
where $S_{n,u}$ is the session state of the $n$-th query session of user $u$, which is equal to the last hidden state of the session-level \ac{RNN}.

As shown in Fig.~\ref{HRNN-QS}, we use the final hidden state of the user-level \ac{RNN} to denote the user state, $U_t=h_{t,u}$, that contains the information about the search behavior from a user's past sessions and thus can be applied in the session-level \ac{RNN}. In \ac{HNQS}, the session-level \ac{RNN} is initialized with a user state as follows:
%
$h_{0,t+1}=\tanh (\textbf{$W$}\cdot U_{t} + \textbf{$b_0$})$.
%
We choose to use only the initialization strategy for session-level \ac{RNN} with user state $U_t$, instead of transporting the user information from $U_t$ throughout the whole session-level \ac{RNN} including initialization, updating and output. The GRU unit has both long and short term memory~\citep{LSTM} and can automatically  transport the user state information within the network, which leads to a better performance when combined with our initialization strategy.

\subsection{Attention-based hierarchical \ac{RNN} for query suggestion}
\label{section:AHNQS}

We assume that submitted queries that trigger subsequent click behavior have a better expression of the user's search intent. We hypothesize that queries in a session should have different weights to reflect the user's information need and employ an attention mechanism on top of the \ac{HNQS} model to capture the user's preference for different queries in a session and then aggregate the representations of informative queries.
\begin{figure}[h]
  \centering
   \includegraphics[width=0.28\textwidth]{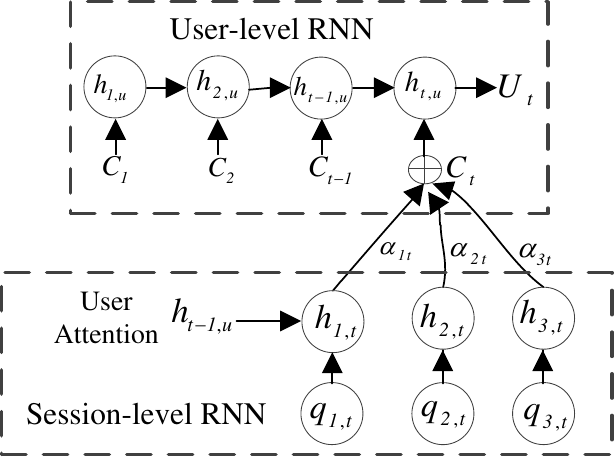}
   \caption{Structure of the \ac{AHNQS} model.}
\label{AHRNN-QS}
\end{figure}
%

Fig.~\ref{AHRNN-QS} shows how we update the user-level \ac{RNN} in \ac{AHNQS} as follows:
$h_{t,u}=\emph{RNN}_\emph{user}(h_{t-1,u},C_{t})$,
where $C_{t}=\sum_{j=1}^M \alpha_{jt}h_{j,t}$ is the attentive representation of the session state, a weighted sum of the hidden states $h_{j,t}$ from the session-level \ac{RNN},
where $\alpha_{jt}$ is the normalized attention score for the $j$-th query in session $\emph{Session}_t$, which is interpreted as the contribution of the query to the preference of the user:
\begin{equation}
\label{equation11}
\alpha_{jt}=\frac{\exp(e_{jt})}{\sum_{k=1}^M \exp(e_{kt})},
\end{equation}
where $e_{jt}=h_{t-1,u}^\mathrm{T}\mathbf{W}_ah_{jt}$ is the initial attention score computed with user state $h_{t-1,u}$ and hidden state $h_{jt}$ in a session-level \ac{RNN},
where the parameters $\mathbf{W}_a$ can be jointly trained with all the other components of \ac{AHNQS}  as the attention mechanism allows the gradient of the loss function to be backpropagated.

Thus, \ac{AHNQS} combines a hierarchical user-session \ac{RNN} and an attention mechanism for query suggestion; the hierarchical structure models the user's short and long-term search behavior, while the attention mechanism captures the user's query preference.


\section{Experiments}
\label{Model}

\noindent%
\textbf{Research questions.}
\begin{inparaenum}[]
\item (\textbf{RQ1}) Do the hierarchical structure and attention mechanism incorporated in \ac{HNQS} and \ac{AHNQS} help to improve the performance of the neural query suggestion model and beat the state-of-the-art method?
\item (\textbf{RQ2}) What is the impact on query suggestion performance of session length, i.e., short vs.\ medium length vs.\ long sessions?

\end{inparaenum}

\smallskip\noindent%
\textbf{Model summary.}
As AHNQS is based on a neural network to capture the dependencies between queries and users, we compare it with the state-of-the-art neural models for query suggestion. As an aside, \citet{RNN2015} incorporate a neural query suggestion model as a feature into a learning to rank approach and thus belongs to the feature engineering approaches we do not compare with.
We consider the following baselines for comparison:
\begin{inparaenum}[(1)]
    \item  ADJ: original co-occurrence based query suggestion method~\citep{Relevant2003};
    \item  NQS: a simple session-based \ac{RNN} method for query suggestion~\citep{2016session-based}, \S\ref{section:NQS}.
In addition, we consider:
    \item  \ac{HNQS}: a hierarchical structure with user-session level \ac{RNN} for query suggestion, \S\ref{section:HNQS};
    \item  \ac{AHNQS}: an attention-based hierarchical \ac{RNN} model for query suggestion, \S\ref{section:AHNQS}.
        %
\end{inparaenum}

%

\smallskip\noindent%
\textbf{Datasets and parameters.}
\label{section:datasetsandparameters}
We use the AOL query log~\citep{Pass:2006} and preprocess the dataset following~\citep{intent2011}.
Queries are separated into sessions by $30$ minutes of inactivity.
We remove queries with less than 20 occurrences and keep sessions with length larger than 5 as well as users with at least 5 sessions to provide sufficient user-session information.
The training set consists all but the last 30 days in the search history; the test set consists of the last 30 days in the log after filtering out queries that do not exist in the training set.
Table~\ref{table1} details the statistics of the dataset used.

\begin{table}[t]
\centering
\caption{Dataset statistics.}
\label{table1}
\begin{tabular}{@{}lrr@{}}
\toprule
    Variable& Training & Test\\
    \midrule
 \# Queries & 1,545,543 & 576,817 \\
 \# Unique queries & 61,641 & 33,519 \\
 \# Sessions & 166,414 & 67,716 \\
 \# Users & 23,308 & 19,255 \\
  Average \# queries per session & 9.28 & 8.52 \\
  Average \# sessions per user & 7.14 & 3.52 \\
  \bottomrule
\end{tabular}
\vspace*{-.5\baselineskip}
\end{table}

We use GRUs as the \ac{RNN} units and optimize the neural models using TOP1 loss function and AdaGrad with momentum for 20 epochs.
The number of hidden units is set to 100 in all cases and we use dropout regularization.
We optimize the hyperparameters by running 100 experiments at randomly selected points of the parameter space.
Optimization is done on a validation set, which is partitioned from the training set with the same procedure as the test set.
We summarize the best performing parameters in Table~\ref{table2}.

\begin{table}[t]
\centering
\caption{Parameters used for each model.}
\label{table2}
\begin{tabular}{lcccc}
\toprule
    Model & Batch & Dropout& Learning rate & Momentum\\
    \midrule
 \ac{NQS} & 50 & 0.5 & 0.01 & 0.0 \\
 \ac{HNQS} & 50 & 0.1 & 0.1\phantom{0} & 0.0 \\
 \ac{AHNQS} & 50 & 0.1 & 0.1\phantom{0} & 0.0 \\
  \bottomrule
\end{tabular}
\end{table}

\smallskip\noindent%
\textbf{Training and evaluation.}
As the lengths of sessions are different and our goal is to capture how a session evolves over time, traditional methods to form batches in natural language processing are not suitable for our query suggestion task.
We thus use parallel mini-batches with the identifications of users and sessions following~\citep{2016session-based}.
We evaluate the models by providing queries in a session one by one and measure the ranking performance of query suggestions with MRR and Recall on the test set.


\section{Results and Discussion}
\label{Results}


\noindent%
\textbf{Overall performance.}
To answer \textbf{RQ1}, we examine the query suggestion performance of the baselines as well as the \ac{HNQS} and \ac{AHNQS} models. Table~\ref{table3} shows the results.
\begin{table}[h]
\captionsetup{justification=justified}
  \centering
  \caption{Performance of query suggestion models. The results by the best baseline and the best performer in each column are underlined and in boldface, respectively. Statistical significance of pairwise differences of HNQS and AHNQS vs.\ the best baseline) is determined by a $t$-test ($^\blacktriangle$/$^\blacktriangledown$ for $\alpha$ = .01).}
\label{table3}
\begin{tabular}{lcc}
\toprule
Model & {Recall@10} & {MRR@10}   \\
\midrule
ADJ & \underline{.7072}\duwen & \underline{.6922}\duwen  \\
NQS & .6444\duwen & .6148\duwen  \\
HNQS & .8138$^\blacktriangle$ &.7874$^\blacktriangle$  \\
AHNQS & \bf{.8618}$^\blacktriangle$ &\bf{.8514}$^\blacktriangle$  \\
\bottomrule
\end{tabular}
\end{table}
ADJ outperforms \ac{NQS}, with 9.74\% and 12.58\% improvements in terms of Recall@10 and MRR@10, respectively. This may be because that the \ac{NQS} model (without knowing about individual users) fails to capture information from the past search history. \ac{HNQS} shows significant improvements over ADJ, with Recall@10 improved 15.07\%  and MRR@10 improved 13.75\%. The hierarchical structure effectively incorporates a given user's previous search behavior and thus improves accuracy.
The best performance is obtained by \ac{AHNQS}, which outperforms ADJ by 21.86\% (Recall@10) and 22.99\% (MRR@10), and \ac{HNQS} by 5.9\% (Recall@10) and 8.13\% (MRR@10). The latter difference indicates that attention can strengthen the model's ability to rank query suggestion candidates effectively.

\begin{figure}[t]
\captionsetup{justification=justified}
        \centering
        \begin{subfigure}[t]{0.24\textwidth}
                \centering
                \includegraphics[clip, trim=0mm 0mm 0mm 1mm, width=0.76\columnwidth]{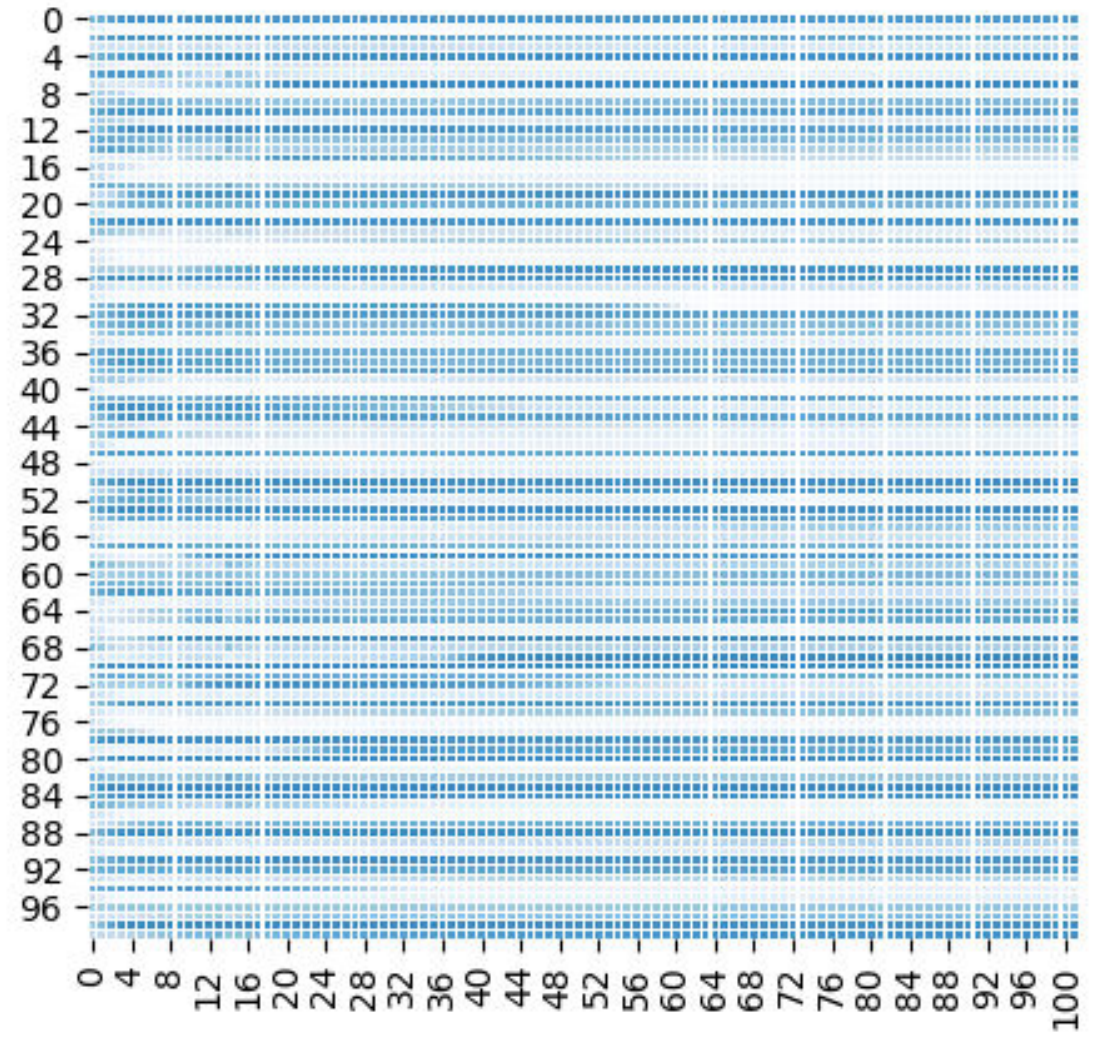}
                \vspace*{-.5\baselineskip}
                \caption{session-level \ac{RNN} in \ac{NQS}.}
                \label{RNN_session}
        \end{subfigure}
~
        \begin{subfigure}[t]{0.24\textwidth}
                \centering
                \includegraphics[clip, trim=0mm 0mm 0mm 1mm, width=0.76\columnwidth]{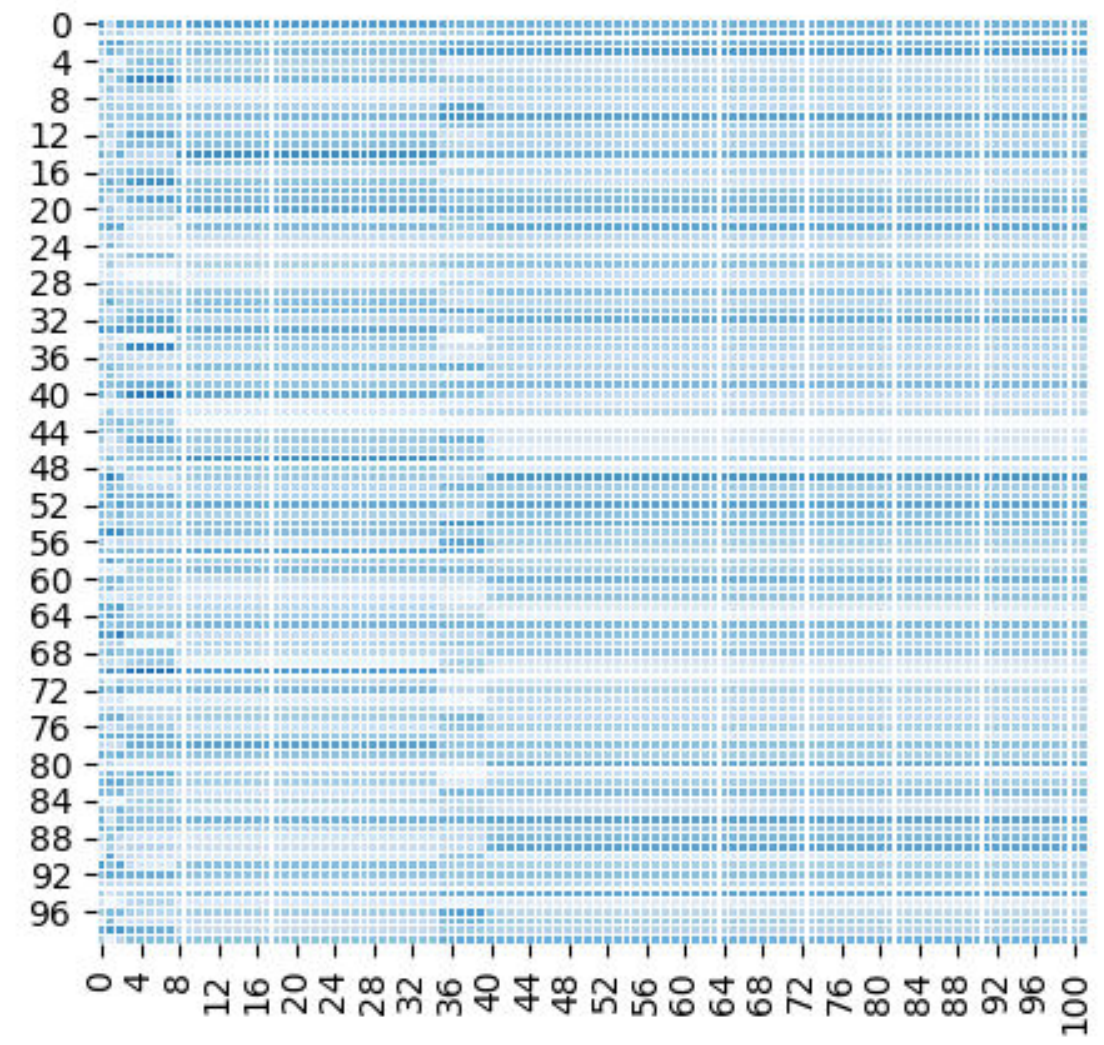}
                \vspace*{-.5\baselineskip}
                \caption{session-level \ac{RNN} in \ac{HNQS}.}
                \label{HRNN_session}
        \end{subfigure}
        \begin{subfigure}[t]{0.24\textwidth}
                \centering
                \includegraphics[clip, trim=0mm 0mm 1mm 0mm, width=0.76\columnwidth]{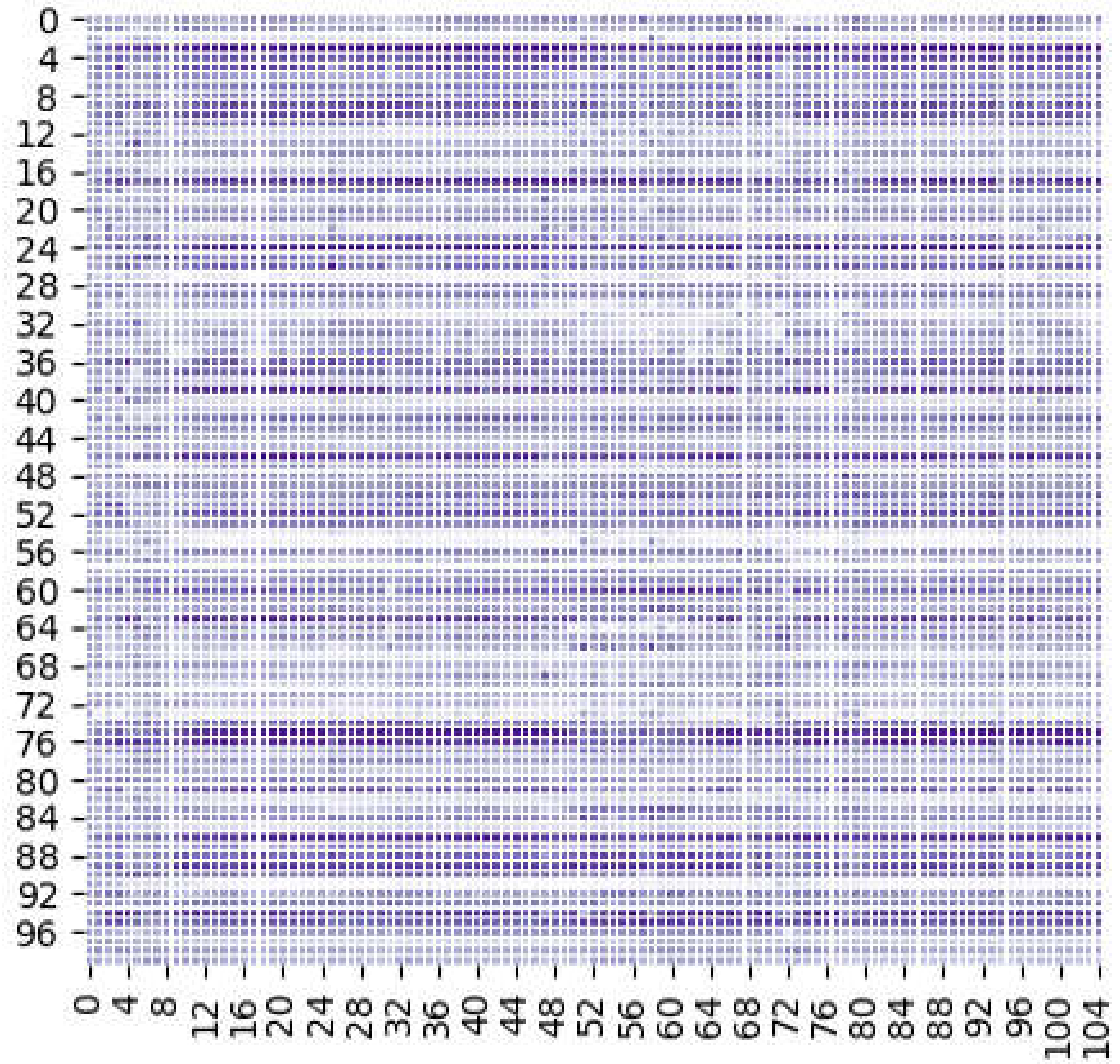}
                \vspace*{-.5\baselineskip}
                \caption{user-level \ac{RNN} in \ac{HNQS}.}
                \label{HRNN_user}
        \end{subfigure}
~
        \begin{subfigure}[t]{0.24\textwidth}
        \centering
                \includegraphics[clip, trim=0mm 0mm 1mm 0mm, width=0.76\columnwidth]{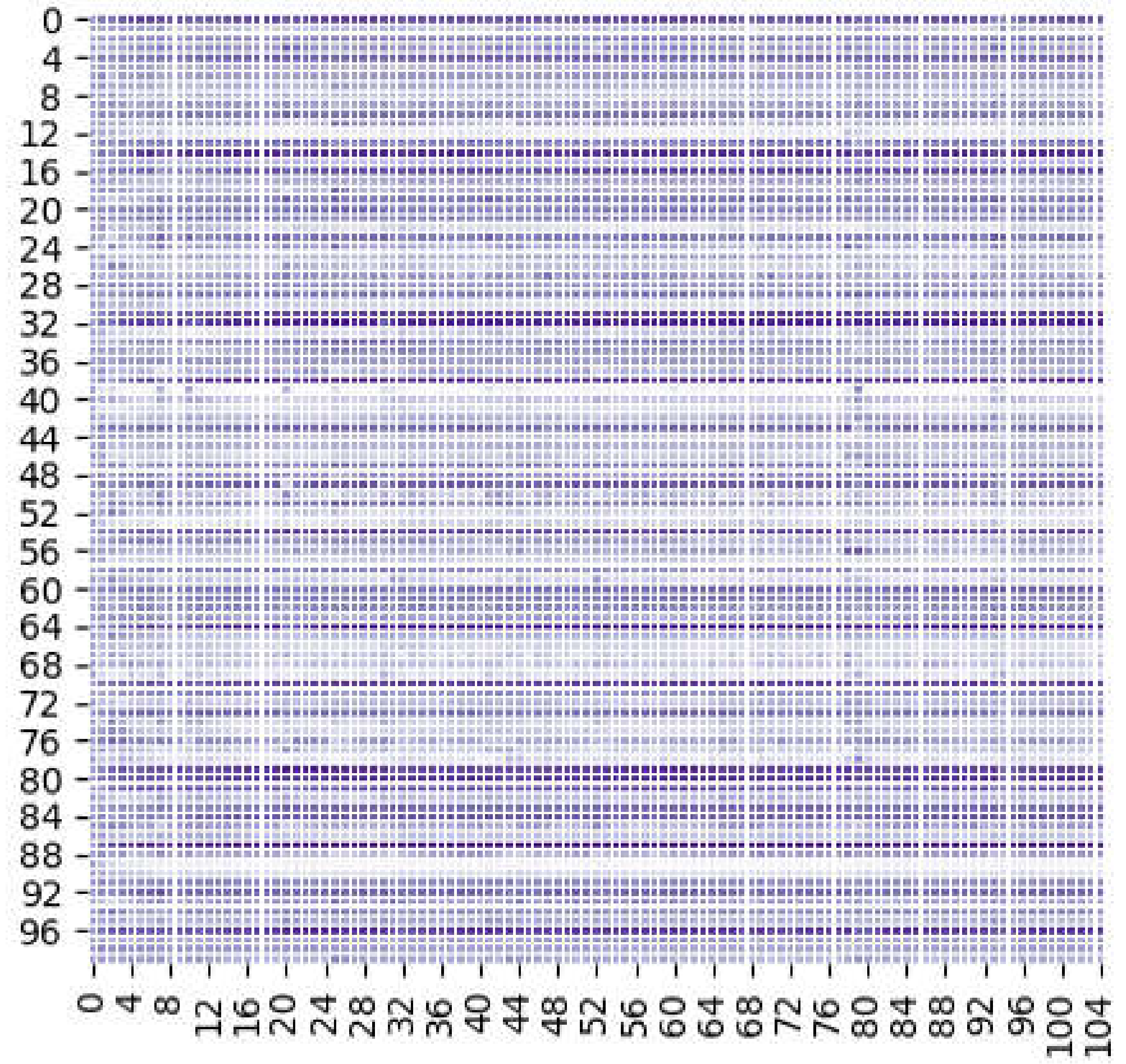}
                \vspace*{-.5\baselineskip}
                \caption{user-level \ac{RNN} in \ac{AHNQS}.}
                \label{AHRNN_user}
        \end{subfigure}
\caption{Visualizing hierarchical structure ((a) and (b)) and attention mechanism ((c) and (d)). The lighter the area in the plot, the more important the information is.}
\label{figure2}
\vspace*{-\baselineskip}
\end{figure}

To determine the impact of the hierarchical structure and attention mechanism, we consider a sample session and user, and compare the hidden states of an \ac{RNN} in \ac{NQS} and \ac{HNQS}, respectively (Fig.~\ref{RNN_session} and~\ref{HRNN_session}), as well as the hidden states of an \ac{RNN} in \ac{HNQS} and \ac{AHNQS}, respectively (Fig.~\ref{HRNN_user} and~\ref{AHRNN_user}). The lighter the area in the plot, the more important the information is.
In Fig.~\ref{RNN_session} and~\ref{HRNN_session}, the session contains 102 queries (x-axis); the number of hidden units is 100 (y-axis). Compared with Fig.~\ref{RNN_session}, we can see that the hierarchical structure modifies the user's search intent especially at the first positions in a session in Fig.~\ref{HRNN_session}.
There is a fluctuation around the 36th to 40th queries in Fig.~\ref{HRNN_session}, which may be due to the fact that we use a GRU unit inside the session-level \ac{RNN} to transport the user state information within the network.

Turning to the attention mechanism (Fig.~\ref{HRNN_user} and~\ref{AHRNN_user}), we select a user with 105 sessions (x-axis) and the number of hidden units in the user-level \ac{RNN} is set to 100 (y-axis).
Compared with Fig.~\ref{HRNN_user}, the user's preference towards different information is more equally distributed inside the user-level \ac{RNN} in Fig.~\ref{AHRNN_user}.
Moreover, going from left to right there are fewer abrupt shifts form high interest (light) to low interest (dark) areas, or vice versa, in Fig.~\ref{AHRNN_user} than in Fig.~\ref{HRNN_user}: the attention mechanism can help to describe a user's long-term search preferences towards different topics.

\begin{figure}[t]
            \centering
                \includegraphics[clip, trim=0mm 1mm 0mm 0mm, width=.78\columnwidth]{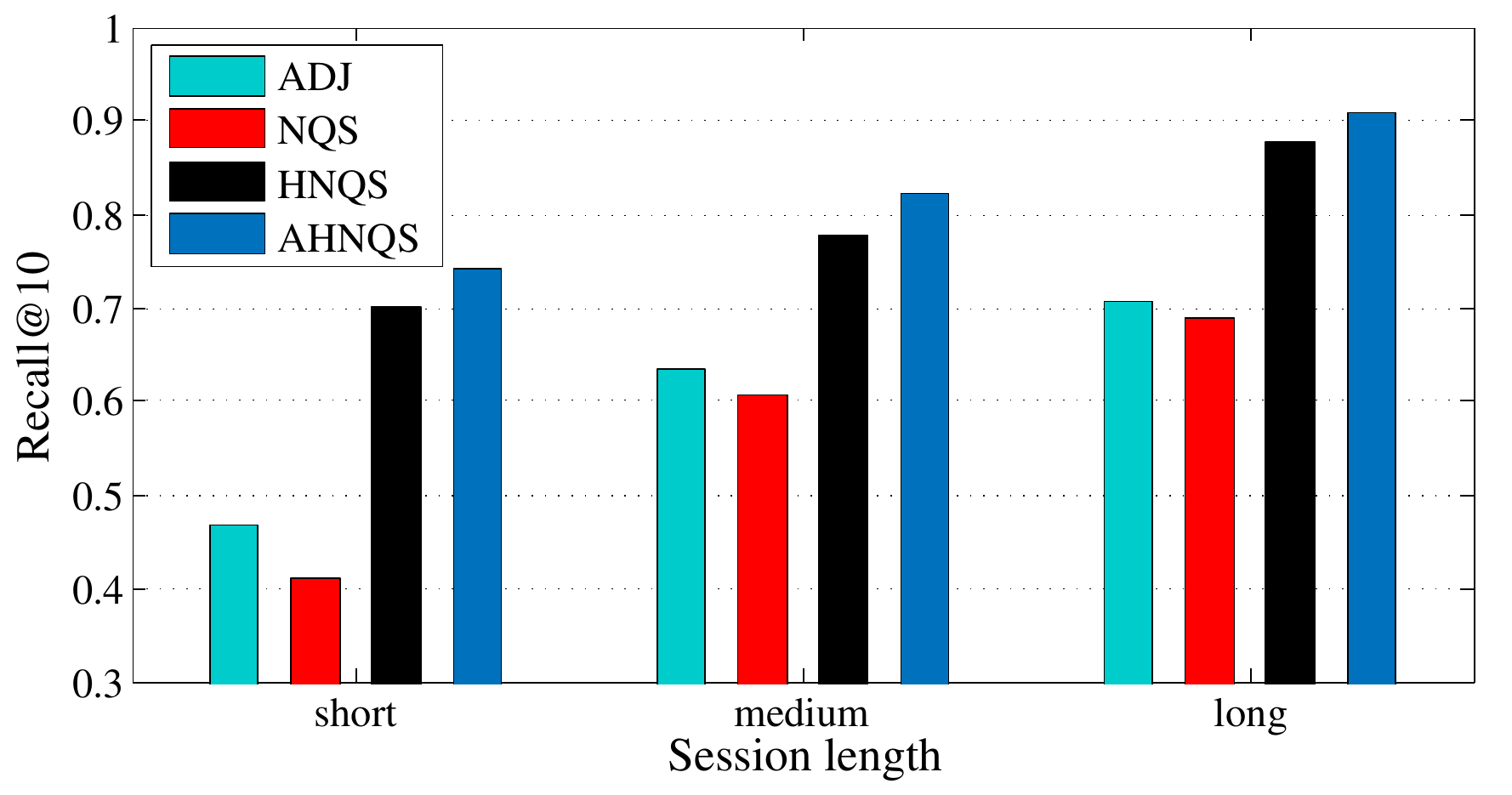}
                \label{session_length_recall}
        \caption{Performance with different session lengths.}
         \vspace*{-1.5\baselineskip}
\label{figure3}
\end{figure}

\smallskip\noindent%
\textbf{Impact of current session length.}
\label{session_length}
For \textbf{RQ2}, we expect the current session length to have an impact on the performance of query suggestion models. We report separate results for short (2 queries), medium (3 or 4 queries) and long current sessions (at least 5 queries) on the test set in Fig.~\ref{figure3}.
%
%

%
As the session length increases, the performance in terms of Recall@10 of all query suggestion models improves and \ac{AHNQS} always achieves the highest scores. As for baselines, ADJ outperforms \ac{NQS}; the margin between them shrinks as the session length increases. For short current sessions, \ac{HNQS} and \ac{AHNQS} yield larger improvements than for medium and long current sessions.
This is due to the fact that when predicting a user's search intent at the first positions of a session, the hierarchical structure within RNN models can provide effective information from a user's past search history and thus can improve the accuracy for query suggestion (compared to ADJ and \ac{NQS}).

For MRR a similar picture emerges (not shown).
\ac{AHNQS} shows bigger improvements over \ac{HNQS}, of 11.23\%, 7.13\% and 8.74\% in terms of MRR@10, for short, medium and long sessions, respectively, vs.\ improvements of 5.88\%, 5.97\% and 3.61\% for Recall@10. This confirms our intuition about attention mechanisms, i.e., that they help to improve precision.


\section{Conclusions and Future Work}
\label{Conclusion}
We have proposed an attention-based hierarchical neural query suggestion model (\ac{AHNQS}) that combines a hierarchical user-session \ac{RNN} with an attention mechanism. 
The hierarchical structure can model both the user's short-term and long-term search behavior effectively, while the attention mechanism captures a user's preference towards certain queries over others. 
The experimental results also confirm the effectiveness of the proposed model for query suggestion across sessions with various lengths.
As to future work, we plan to evaluate our model on other datasets so as to verify its robustness. 
We also want to investigate the performance of \ac{AHNQS} when combining semantic similarity with the hierarchical structure, i.e., different encoding methods for input queries.


\smallskip
\begin{spacing}{1}
\noindent
\textbf{Acknowledgments.}
This research was supported by
the National Natural Science Foundation of China under No. 61702526,
the National Advanced Research Project under No. 6141B0801010b,
Ahold Delhaize,
Amsterdam Data Science,
the Bloomberg Research Grant program,
the China Scholarship Council,
the Criteo Faculty Research Award program,
Elsevier,
the European Community's Seventh Framework Programme (FP7/2007-2013) under
grant agreement nr 312827 (VOX-Pol),
the Google Faculty Research Awards program,
the Microsoft Research Ph.D.\ program,
the Netherlands Institute for Sound and Vision,
the Netherlands Organisation for Scientific Research (NWO)
under pro\-ject nrs
CI-14-25, 
652.\-002.\-001, 
612.\-001.\-551, 
652.\-001.\-003, 
and
Yandex.
All content represents the opinion of the authors, which is not necessarily shared or endorsed by their respective employers and/or sponsors.
\end{spacing}

%

\vspace*{-.1\baselineskip}
\bibliographystyle{ACM-Reference-Format}
\bibliography{sigproc}

\end{document}